\documentclass[aps,pra,floatfix,reprint,showpacs,10pt]{revtex4-1}

\usepackage{graphicx}
\usepackage{dcolumn}
\usepackage{bm}
\usepackage{amssymb}
\usepackage{rotating} 

\begin{document}


\title{Significant Conditions on the Two-electron Reduced Density Matrix from the Constructive Solution of $N$-representability}

\author{David A. Mazziotti}

\email{damazz@uchicago.edu}
\affiliation{Department of Chemistry and The James Franck Institute, The University of Chicago, Chicago, IL 60637}%

\date{Submitted May 4, 2012; Published {\em Phys. Rev. A} {\bf 85}, 062507 (2012)}


\begin{abstract}

We recently presented a constructive solution to the
$N$-representability problem of the two-electron reduced density
matrix (2-RDM)---a systematic approach to constructing complete
conditions to ensure that the 2-RDM represents a realistic
$N$-electron quantum system [D. A. Mazziotti, Phys. Rev. Lett. {\bf
108}, 263002 (2012)]. In this paper we provide additional details
and derive further $N$-representability conditions on the 2-RDM that
follow from the constructive solution. The resulting conditions can
be classified into a hierarchy of constraints, known as the
$(2,q)$-positivity conditions where the $q$ indicates their
derivation from the nonnegativity of $q$-body operators.  In
addition to the known T1 and T2 conditions, we derive a new class of
(2,3)-positivity conditions. We also derive 3 classes of
(2,4)-positivity conditions, 6 classes of (2,5)-positivity
conditions, and 24 classes of (2,6)-positivity conditions.  The
constraints obtained can be divided into two general types: (i) {\em
lifting conditions}, that is conditions which arise from lifting
lower $(2,q)$-positivity conditions to higher $(2,q+1)$-positivity
conditions and (ii) {\em pure conditions}, that is conditions which
cannot be derived from a simple lifting of the lower conditions. All
of the lifting conditions and the pure $(2,q)$-positivity conditions
for $q>3$ require tensor decompositions of the coefficients in the
model Hamiltonians.  Subsets of the new $N$-representability
conditions can be employed with the previously known conditions to
achieve polynomially scaling calculations of ground-state energies
and 2-RDMs of many-electron quantum systems even in the presence of
strong electron correlation.

\end{abstract}


\maketitle

\section{Introduction}

Because electrons are indistinguishable with pairwise Coulomb
interactions, the energies and properties of many-electron atoms and
molecules can be evaluated from a knowledge of the two-electron
reduced density matrix (2-RDM)~\cite{RDM,CY00,M12a}.  Minimizing the
ground-state energy as a functional of the 2-RDM, however, requires
non-trivial constraints on the 2-RDM to ensure that it represents an
$N$-electron system ($N$-representability conditions)~\cite{RDM,
CY00,M12a,C63,GP64,K67,H78,E78,P78,EJ00,ME01,N01,M02,M04,P04,C06,M05,
M06,GM07,E07,A09,SI10,M11,BP12}.  While advances in theory and
computation enabled the accurate variational calculation of the
2-RDM for a variety of strongly correlated systems in chemistry and
physics from polyaromatic hydrocarbons~\cite{GM08,PGG11} to quantum
dots~\cite{RM09}, the known $N$-representability conditions for the
2-RDM, albeit rigorous, remained incomplete. Recently, we presented
a constructive solution to the $N$-representability problem---a
systematic approach to constructing complete $N$-representability
conditions on the two-electron reduced density matrix (2-RDM)---as
well as examples of new $N$-representability conditions~\cite{M12b}.
In the present paper we present additional details as well as
further conditions on the 2-RDM that follow from the constructive
solution.

The advantage of reduced variables such as the 2-RDM and the
one-electron density is that, unlike the wavefunction expanded in
terms of determinants, their degrees of freedom grow {\em
polynomially} with the size of the quantum system~\cite{M12a} even
when the electrons are strongly correlated~\cite{ZK05,MA03}. Direct
calculation of the reduced variables, however, requires that they
and their functionals be consistent with a realistic $N$-electron
quantum system; in other words, the reduced variables and
functionals must be representable by the integration of an
$N$-electron density matrix.  Such consistency relations are known
as the {\em $N$-representability conditions}~\cite{RDM,
CY00,M12a,C63,GP64,K67,H78,E78,P78,EJ00,ME01,
N01,M02,M04,P04,C06,M05,M06,E07,A09,SI10,M11,BP12}.  These
conditions are particularly important to 2-RDM methods where they
enable the direct calculation of the 2-RDM without the wavefunction,
but they are also implicit in the design of realistic approximations
to the density functional in density functional theory~\cite{PY94,
CH08}.

Minimizing the many-electron energy as a functional of the 2-RDM
{\em without} $N$-representability conditions produces an energy
that is {\em much lower} than the exact ground-state energy of the
quantum system.  The energy is too low because both the energy and
the computed 2-RDM are not realistic---they are not
$N$-representable. In the early 1960s the search for the set of
necessary and sufficient $N$-representability conditions became
known as the $N$-representability problem~\cite{C63}.  Three
important constraints, known as the D, Q, and G (or 2-positivity)
conditions, were developed by Coleman~\cite{C63} and Garrod and
Percus~\cite{GP64}. The D, Q, and G conditions restrict the
probability distributions of two electrons, two holes (where a {\em
hole} is the absence of an electron), and an electron-hole pair to
be nonnegative.  Each condition can be expressed in the form of
constraining a matrix to be positive semidefinite.  A matrix is {\em
positive semidefinite} if and only if its eigenvalues are
nonnegative.

In 1978 Erdahl~\cite{E78} discovered two additional semidefinite
constraints on the 2-RDM known as the T1 and T2 (or partial
3-positivity) conditions~\cite{P04,M06,P07,M05}, which are derivable
from the nonnegativity of the three-electron probability
distributions.  Finally, Weinhold and Wilson~\cite{WW67}, Yoseloff
and Kuhn~\cite{YK69}, McRae and Davidson~\cite{MD72}, and
Erdahl~\cite{E78} derived necessary conditions on the {\em diagonal}
part of the 2-RDM.  These diagonal conditions were shown, in the
context of the Boole optimization problem~\cite{Cuts}, to be part of
a complete set of {\em classical $N$-representability conditions} on
the two-electron reduced density function which is the diagonal part
of the 2-RDM in a coordinate representation~\cite{KM08}.  Despite
the solution of the classical problem, the complete set of quantum
$N$-representability conditions remained unknown except for the D,
Q, G, T1, and generalized T2 conditions as well as unitary
transformations of the classical $N$-representability conditions. In
2001 Mazziotti and Erdahl~\cite{ME01} presented a systematic
generalization of these constraints known as the {\em $p$-positivity
conditions} and in 2002 Mazziotti~\cite{M02,HM05} introduced the
{\em lifting conditions}; however, except for the conditions given
above, the $p$-positivity conditions and the lifting conditions
depend upon not only the 2-RDM but also higher-particle RDMs.

The constructive solution to the $N$-representability problem
provides a systematic approach to building complete
$N$-representability conditions on the two-electron reduced density
matrix (2-RDM)~\cite{M12b}.  While an example of the new conditions
was given previously, in the present paper we present further
$N$-representability conditions on the 2-RDM that follow from the
constructive solution.  The conditions are in the form of a set of
model Hamiltonians with pairwise interactions whose trace against
the 2-RDM must be nonnegative.  The resulting conditions can be
classified into an increasing hierarchy of constraints, known as the
$(2,q)$-positivity conditions where the first number $p$ in the name
indicates the highest $p$-RDM required to evaluate the condition
(the 2-RDM in our case) and the second number $q$ indicates the
highest $q$-particle reduced density operators ($q$-RDOs) canceled
by nonnegative linear combinations in the derivation of the
condition. The $(p,p)$-positivity conditions are equivalent to the
$p$-positivity conditions introduced earlier in
Refs.~\cite{EJ00,ME01,M02}.  We will use the two conventions in
nomenclature interchangeably.

In addition to the previously known T1 and T2 conditions~\cite{E78,
P04,M06,P07,M05}, we derive a new class of (2,3)-positivity
conditions. We also derive 3 classes of (2,4)-positivity conditions,
6 classes of (2,5)-positivity conditions, and 24 classes of
(2,6)-positivity conditions.  The conditions obtained can be divided
into two general types: (i) {\em lifting conditions}, that is
conditions which arise from lifting lower $(2,q)$-positivity
conditions to higher $(2,q+1)$-positivity conditions and (ii) {\em
pure conditions}, that is conditions which cannot be derived from a
simple lifting of the lower conditions. All of the lifting
conditions and the pure $(2,q)$-positivity conditions for $q>3$
require that the expansion coefficients in the model Hamiltonians be
{\em tensor decomposed}. Subsets of the $N$-representability
conditions can be employed with previously known conditions for
polynomially scaling calculations of ground-state energies and
2-RDMs of many-electron quantum systems in chemistry and physics.

\section{Theory}

After the constructive solution of $N$-representability is reviewed
in section~\ref{sec:sol}, it is employed in sections~\ref{sec:kcon}
and~\ref{sec:ncon} to derive known and new $N$-representability
conditions, respectively.  The new constraints are organized into
sections on (2,3)-, (2,4)-, (2,5)-, and (2,6)-positivity conditions.
Two algorithms for implementing the conditions in a variational
2-RDM calculation are briefly discussed in section~\ref{sec:sd}.

\subsection{Constructive solution}

\label{sec:sol}

The energy of an $N$-electron quantum system in a stationary state
can be computed from the Hamiltonian traced against the state's
density matrix
\begin{equation}
\label{eq:EN} E = {\rm Tr}({\hat H} \, {}^{N} D) ,
\end{equation}
where the Hamiltonian operator is expressible in second quantization
as
\begin{equation}
{\hat H} = \sum_{ijkl}{ {}^{2} K^{ij}_{kl} {\hat a}^{\dagger}_{i}
{\hat a}^{\dagger}_{j} {\hat a}_{l} {\hat a}_{k} }
\end{equation}
in which the matrix ${}^{2} K$ is the reduced Hamiltonian operator
in a finite one-electron basis set~\cite{M98} and the indices label
the members (orbitals) of the basis set. Because electrons are
indistinguishable with pairwise interactions, the energy can also be
universally written as a linear functional of only the 2-RDM
\begin{equation}
\label{eq:E2} E = {\rm Tr}({\hat H} \, {}^{2} D) ,
\end{equation}
where the 2-RDM can be formally defined from integration of the
$N$-electron density matrix over all electrons save two
\begin{equation}
\label{eq:D2} {}^{2} D = \frac{N(N-1)}{2} \int{ {}^{N} D \, d3 \dots
dN } .
\end{equation}
The expression of the energy as a functional of the 2-RDM suggests
the tantalizing possibility of computing the ground-state energy of
any electronic system as a functional of only the 2-RDM~\cite{M55,
RDM,CY00}.  Early calculations by Coleman~\cite{C63},
Tredgold~\cite{T57}, and others, however, showed that minimization
of the energy as a 2-RDM functional produces unphysically low
energies without additional constraints on the 2-RDM to ensure that
it represents an $N$-electron density matrix. In 1963 Coleman called
these constraints the $N$-representability conditions~\cite{C63}.

Building upon work by Garrod and Percus~\cite{GP64}, Kummer in 1967
showed by the bipolar theorem~\cite{R71} that there exists a convex
set (cone) of two-body operators $\{ {}^{2} {\hat O_{i}} \}$ whose
trace against a potential 2-RDM will be nonnegative
\begin{equation}
{\rm Tr}({}^{2} {\hat O} \, {}^{2} D) \ge 0
\end{equation}
if and only if the 2-RDM is $N$-representable~\cite{K67}. Hence, the
set of two-body operators $\{ {}^{2} {\hat O_{i}} \}$ defines the
set $P^{2}_{N}$ of $N$-representable 2-RDMs.  We say that the set
$\{ {}^{2} {\hat O_{i}} \}$ is the polar of $P^{2}_{N}$ and denote
it as ${P^{2}_{N}}^{*}$.  Characterizing the set $P^{2}_{N}$ of
$N$-representable 2-RDMs, therefore, would be complete if we could
characterize its polar set ${P^{2}_{N}}^{*}$. Kummer's original
result demonstrates the existence of the set ${P^{2}_{N}}^{*}$, but
it does not provide a prescription for constructing it.

Recently, a constructive solution to the $N$-representability
problem has been derived through the complete characterization of
the polar set ${P^{2}_{N}}^{*}$~\cite{M12b}.  In Ref.~\cite{M12b} it
is proven that the second-quantized representation of the operators
$\{ {}^{2} {\hat O_{i}} \}$ in ${P^{2}_{N}}^{*}$ can be explicitly
constructed as follows
\begin{equation}
\label{eq:O2} {}^{2} {\hat O} = \sum_{i}{ w_{i} {\hat C}_{i} {\hat
C}_{i}^{\dagger} }
\end{equation}
where ${\hat C}_{i}$ are polynomials in the creation and/or
annihilation operators of degree less than or equal to $r$ (the rank
of the one-electron basis set) and $w_{i}$ are nonnegative integer
weights. The proof relies on the fact that ${P^{2}_{N}}^{*}$ is {\em
contained within} the set ${P^{r}_{N}}^{*}$ of operators of degree
$\le 2r$ whose trace against an $N$-electron density matrix must be
nonnegative. Because the extreme elements (rays) of the convex cone
${P^{r}_{N}}^{*}$ are readily expressed as~\cite{H02}
\begin{equation}
{\hat C}_{i} {\hat C}_{i}^{\dagger} ,
\end{equation}
the extreme elements (rays) of ${P^{2}_{N}}^{*}$ can be constructed
from the {\em conic combinations} (or nonnegative linear
combinations) given in Eq.~(\ref{eq:O2}).  The conic combinations,
if divided by $\sum_{i}{w_{i}}$, can be interpreted as {\em convex
combinations}. Conic combinations are contained in ${P^{2}_{N}}^{*}$
if and only if they cancel all three- and higher-body operators,
that is polynomials in creation and annihilation operators of degree
greater than of equal to 6.

\subsection{Practical implementation}

\label{sec:sd}

Before developing known and new $N$-representability conditions in
sections~\ref{sec:kcon} and~\ref{sec:ncon} respectively, in this
section we briefly indicate their practical applications by
sketching two algorithms for computing the ground-state 2-RDM.
Minimizing the ground-state energy as a function of the 2-RDM
constrained by these conditions can be formulated as a linear
program
\begin{eqnarray}
\label{eq:Ex} {\rm minimize~~}  E & = & {\rm Tr}({\hat H} \, {}^{2}
D) \\ {\rm such~that~~} {\rm Tr}({\hat O}_{j} \, {}^{2} D ) & \ge &
0~~~{\rm for~all}~j, \label{eq:lin}
\end{eqnarray}
in which the necessary set of operators (model Hamiltonians) ${\hat
O}_{j}$, defining the boundary of the convex set of 2-RDMs, must be
determined iteratively.  Given an initial set of model-Hamiltonian
constraints that bound the minimum energy, the three key steps of
the algorithm are: (i) solving the linear program for the optimal
2-RDM, (ii) updating the set of model-Hamiltonian constraints in the
linear program, and (iii) repeating steps (i) and (ii) until the
2-RDM is nonnegative in its trace with all model Hamiltonians
explored in step (ii). In the second step, the trace of each model
Hamiltonian with the 2-RDM is minimized by optimizing the
Hamiltonian's parameters (expansion coefficients), and if the final
trace is negative, the model Hamiltonian with its optimized
parameters is added to the constraints in Eq.~(\ref{eq:lin}).  In
practice, only a subset of model Hamiltonians from the constructive
solution is employed.

Some of the $N$-representability constraints can be collected
together as a single semidefinite constraint on the 2-RDM.  The
generalization of a linear program to include semidefinite
constraints is known as a {\em semidefinite program}, and the
solution of such a program is called {\em semidefinite
programming}~\cite{VB96,HSV00}. Efficient large-scale semidefinite
programming algorithms have been developed for the variational
calculation of the 2-RDM~\cite{M04,P04,C06,FNY07,M07,A09,M11,
BP12,E79}.  While the model Hamiltonians corresponding to previously
known $N$-representability conditions in section~\ref{sec:kcon} can
be expressed as semidefinite constraints, the model Hamiltonians
corresponding to the new conditions in section~\ref{sec:ncon}, which
use tensor decompositions of the expansion coefficients in the
${\hat C}_{i}$ operators, cannot be written as traditional
semidefinite constraints.  In practice, however, we can add these
non-standard constraints to a semidefinite program containing the
standard semidefinite constraints by the three-step iterative
procedure discussed above for the linear program.  A main advantage
of this second algorithm is that a large number of model
Hamiltonians can be included by a single semidefinite constraint. A
similar algorithm, to which we refer for further details, was
proposed in Ref.~\cite{JSM07} for imposing the T2 condition by
recursively generated linear inequalities.

\subsection{Known conditions}

\label{sec:kcon}


All previously known $N$-representability conditions are generated
by the constructive solution.  The most important representability
conditions on the 2-RDM, derived by Coleman~\cite{C63} and Garrod
and Percus~\cite{GP64}, are the D, Q, and G conditions---also, known
as the 2-positivity conditions~\cite{ME01}.  These conditions
restrict the two-particle RDM $^{2} D$, the two-hole RDM $^{2} Q$,
and the particle-hole RDM $^{2} G$ to be positive semidefinite, that
is
\begin{eqnarray}
^{2} D & \succeq & 0 \\
^{2} Q & \succeq & 0 \\
^{2} G & \succeq & 0 ,
\end{eqnarray}
where the elements of the RDMs are given by
\begin{eqnarray}
^{2} D^{ij}_{kl} = \langle \Psi | {\hat a}^{\dagger}_{i}
{\hat a}^{\dagger}_{j} {\hat a}_{l} {\hat a}_{k} | \Psi \rangle \\
^{2} Q^{ij}_{kl} = \langle \Psi | {\hat a}_{i} {\hat
a}_{j} {\hat a}^{\dagger}_{l} {\hat a}^{\dagger}_{k} | \Psi \rangle \\
^{2} G^{ij}_{kl} = \langle \Psi | {\hat a}^{\dagger}_{i} {\hat
a}_{j} {\hat a}^{\dagger}_{l} {\hat a}_{k} | \Psi \rangle
\end{eqnarray}
and $M\succeq 0$ indicates that the matrix $M$ is constrained to be
positive semidefinite.  Physically, these conditions correspond to
constraining the probability distributions of two particles, two
holes, as well as one particle and one hole to be nonnegative. The
2-positivity conditions are generated from the constructive solution
by restricting the following three two-body operators from
Eq.~(\ref{eq:O2}) to be nonnegative for all coefficients $b_{ij}$
\begin{eqnarray}
{}^{2} {\hat O}_{D} & = & {\hat C}_{D} {\hat C}_{D}^{\dagger} \label{eq:OD} \\
{}^{2} {\hat O}_{Q} & = & {\hat C}_{Q} {\hat C}_{Q}^{\dagger} \label{eq:OQ} \\
{}^{2} {\hat O}_{G} & = & {\hat C}_{G} {\hat C}_{G}^{\dagger}
\label{eq:OG}
\end{eqnarray}
where the ${\hat C}_{D}$, ${\hat C}_{Q}$, and ${\hat C}_{G}$ cover
all polynomials in creation and annihilation operators of degree two
\begin{eqnarray}
{\hat C}_{D} & = & \sum_{ij}{ b_{ij} {\hat a}^{\dagger}_{i}
{\hat a}^{\dagger}_{j} } \\
{\hat C}_{Q} & = & \sum_{ij}{ b_{ij} {\hat a}_{i}
{\hat a}_{j} } \\
{\hat C}_{G} & = & \sum_{ij}{ b_{ij} {\hat a}^{\dagger}_{i} {\hat
a}_{j} } .
\end{eqnarray}
Note that conic combinations are not present in these conditions
because when the ${\hat C_{i}}$ operators are of degree 2, the
expectation values of the ${\hat O_{i}}$ operators only involve the
2-RDM~\cite{M12b}.

The other previously known $N$-representability conditions---the T1
and T2 conditions~\cite{E78, P04,M06,P07,M05}---are part of the
(2,3)-conditions that follow from the constructive solution.  These
semidefinite conditions on the 2-RDM are obtainable from conic
combinations of three-particle metric matrices that cancel their
dependence on the 3-RDM~\cite{M06,M05}
\begin{eqnarray}
T1 = {}^{3} D + {}^{3} Q & \succeq & 0 \\
T2 = {}^{3} E + {}^{3} F & \succeq & 0 \label{eq:T2}.
\end{eqnarray}
where in second quantization the matrix elements of these metric
matrices are definable as
\begin{eqnarray}
{}^{3} D^{ijk}_{pqs} & = & \langle \Psi | {\hat a}^{\dagger}_{i}
{\hat a}^{\dagger}_{j} {\hat a}^{\dagger}_{k} {\hat a}_{s} {\hat
a}_{q} {\hat a}_{p} | \Psi \rangle \\
{}^{3} E^{ijk}_{pqs} & = & \langle \Psi | {\hat a}^{\dagger}_{i}
{\hat a}^{\dagger}_{j} {\hat a}_{k} {\hat a}^{\dagger}_{s} {\hat
a}_{q} {\hat a}_{p} | \Psi \rangle \\
{}^{3} F^{ijk}_{pqs} & = & \langle \Psi | {\hat a}_{p} {\hat a}_{q}
{\hat a}^{\dagger}_{s} {\hat a}_{k} {\hat a}^{\dagger}_{j}
{\hat a}^{\dagger}_{i} | \Psi \rangle \\
{}^{3} Q^{ijk}_{pqs} & = & \langle \Psi | {\hat a}_{p} {\hat a}_{q}
{\hat a}_{s} {\hat a}^{\dagger}_{k} {\hat a}^{\dagger}_{j} {\hat
a}^{\dagger}_{i} | \Psi \rangle
\end{eqnarray}
The four metric matrices ${}^{3} D$, ${}^{3} E$, ${}^{3} F$, and
${}^{3} Q$ correspond to the probability distributions for three
particles, two particles and a hole, one particle and two holes, and
three holes, respectively~\cite{ME01,M02,M06}.  Restricting the
${}^{3} D$, ${}^{3} E$, ${}^{3} F$, and ${}^{3} Q$ matrices to be
positive semidefinite generates the 3-positivity
conditions~\cite{ME01,M06} which depend on the 3-RDM.  While the T1
and T2 conditions are a subset of the 3-positivity conditions, they
depend only upon the 2-RDM because the 3-particle parts of ${}^{3}
D$ and ${}^{3} Q$ (and ${}^{3} E$ and ${}^{3} F$) cancel upon
addition~\cite{M06,M05}.  For example, the matrix elements of $T1$
are given by
\begin{equation}
T1^{ijk}_{pqs} = 6 \, {}^{3} I^{ijk}_{pqs} - 18 \, ^{1} D^{i}_{p}
\wedge {}^{2} I^{jk}_{qs} + 9 \, {}^{2} D^{ij}_{pq} \wedge
{}^{1}I^{k}_{s} ,
\end{equation}
where $^{p} I$ is the $p$-particle identity matrix and $\wedge$
denotes the Grassmann wedge product~\cite{S71,M98}.

While the T1 condition is unique, three distinct forms of the T2
condition can be generated from rearranging the second-quantized
operators in the definition of the ${}^{3} F$ metric matrix relative
to those in the ${}^{3} E$ metric matrix~\cite{M06}.  Consider the
two variants of the ${}^{3} F$ matrix with the following matrix
elements:
\begin{eqnarray}
{}^{3} {\bar F}^{ijk}_{pqs} & = & \langle \Psi | {\hat a}_{p} {\hat
a}^{\dagger}_{s} {\hat a}_{q} {\hat a}^{\dagger}_{j}
{\hat a}_{k} {\hat a}^{\dagger}_{i} | \Psi \rangle \\
{}^{3} {\tilde F}^{ijk}_{pqs} & = & \langle \Psi | {\hat
a}^{\dagger}_{s} {\hat a}_{p} {\hat a}_{q} {\hat a}^{\dagger}_{j}
{\hat a}^{\dagger}_{i} {\hat a}_{k} | \Psi \rangle .
\end{eqnarray}
The 3-positivity condition ${}^{3} F \succeq 0$ implies both ${}^{3}
{\bar F} \succeq 0$ and ${}^{3} {\tilde F} \succeq 0$ because
reordering the creation and annihilation operators does not change
the vector space covered by the metric matrix.  Changing the
ordering of the second-quantized operators in the ${}^{3} F$ matrix
relative to those in the ${}^{3} E$ matrix, however, does generate
two additional T2 conditions
\begin{eqnarray}
{\bar T2}   & = & {}^{3} E + {}^{3} {\bar F} \succeq 0 \label{eq:T2b}\\
{\tilde T2} & = & {}^{3} E + {}^{3} {\tilde F} \succeq 0
\label{eq:T2t}.
\end{eqnarray}
It was the ${\tilde T2}$ form of the T2 condition that was
originally implemented by Zhao {\em et al.}~\cite{P04} and
Mazziotti~\cite{M05,M06}.

The three T2 conditions are generated in the constructive solution
by keeping the following two-body operators from Eq.~(\ref{eq:O2})
nonnegative
\begin{eqnarray}
^{2} {\hat O}_{T2} & = &  {\hat C}_{E} \,  {\hat
C}_{E}^{\dagger} +  {\hat C}_{F}  {\hat C}_{F}^{\dagger} \\
^{2} {\hat O}_{\bar T2} & = &  {\hat C}_{E} \,  {\hat
C}_{E}^{\dagger} +  {\hat C}_{\bar F}  {\hat C}_{\bar F}^{\dagger} \\
^{2} {\hat O}_{\tilde T2} & = &  {\hat C}_{E} \,  {\hat
C}_{E}^{\dagger} +  {\hat C}_{\tilde F}  {\hat C}_{\tilde
F}^{\dagger}
\end{eqnarray}
where
\begin{eqnarray}
{\hat C}_{E} & = & \sum_{ijk}{ b_{ijk} {\hat a}^{\dagger}_{i} {\hat
a}^{\dagger}_{j} {\hat a}_{k} } \\
{\hat C}_{F} & = & \sum_{ijk}{ b^{*}_{ijk} {\hat a}_{i} {\hat a}_{j}
{\hat a}^{\dagger}_{k} } \\
{\hat C}_{\bar F} & = & \sum_{ijk}{ b^{*}_{ijk} {\hat a}_{i} {\hat
a}^{\dagger}_{k} {\hat a}_{j} } \\
{\hat C}_{\tilde F} & = & \sum_{ijk}{ b^{*}_{ijk} {\hat
a}^{\dagger}_{k} {\hat a}_{i} {\hat a}_{j} } .
\end{eqnarray}
The three T2 conditions can be combined into a single generalized T2
condition as shown in Refs.~\cite{M06,P07}.  The T1 condition is
also produced in the constructive solution by keeping the following
two-body operator from Eq.~(\ref{eq:O2}) nonnegative
\begin{equation}
^{2} {\hat O}_{T1} = {\hat C}_{D} \,  {\hat C}_{D}^{\dagger} + {\hat
C}_{Q}  {\hat C}_{Q}^{\dagger}
\end{equation}
where
\begin{eqnarray}
{\hat C}_{D} & = & \sum_{ijk}{ b_{ijk} {\hat a}^{\dagger}_{i} {\hat
a}^{\dagger}_{j} {\hat a}^{\dagger}_{k} } \label{eq:d3} \\
{\hat C}_{Q} & = & \sum_{ijk}{ b^{*}_{ijk} {\hat a}_{i} {\hat a}_{j}
{\hat a}_{k} } .
\end{eqnarray}
Because the second-quantized operators in ${\hat C}_{D}$ and ${\hat
C}_{Q}$ are anticommutative, there is only one T1 condition.  Unlike
the D, Q, and G conditions, both T1 and T2 conditions arise from the
conic combination of a pair of 3-positive operators that cancels
their dependence on the 3-RDM.

\subsection{New conditions}

\label{sec:ncon}

The constructive solution also produces new $N$-representability
conditions on the 2-RDM~\cite{M12b}.  In this section we will
discuss the further conditions on the 2-RDM that emerge from conic
combinations of three-, four-, five-, and six-particle operators in
Eq.~(\ref{eq:O2}), which we denote as (2,3)-, (2,4)-, (2,5)-, and
(2,6)-positivity conditions, respectively.  All of the new
$N$-representability conditions require a nonlinear factorization of
the expansion coefficients to cancel the higher-particle operators.

\subsubsection{(2,3)-positivity conditions}



\begin{table*}[t!]

\caption{The (2,3)-positivity conditions can be derived from conic
(linear nonnegative) combinations of the (3,3)-positivity conditions
that cancel the 3-particle operators.}

\label{t:23}

\begin{ruledtabular}
\begin{tabular}{cccc}

Class & Type & Representative Condition & ${\hat C}$ Definition \\

\hline

1 & Lifted (2,2) & ${\rm Tr}( ({\hat O}(k,i,j) + {\hat O}({\bar
k},i,j)) \, {}^{2} D)
\ge 0$ & Eq.~(\ref{eq:23P2}) \\

2 & Pure (2,3) & ${\rm Tr}( ({\hat O}(i,j,k) + {\hat O}({\bar
i},{\bar j},{\bar k})) \, {}^{2} D)
\ge 0$ & Eq.~(\ref{eq:d3}) \\

\end{tabular}
\end{ruledtabular}

\end{table*}

In addition to the T1 and T2 conditions there exists a second class
of (2,3)-positivity conditions that can be generated from lifting
the 2-positivity conditions to the three-particle space and then
canceling the three-particle operators.  Consider the pair of
three-body operators
\begin{eqnarray}
{\hat O}(i,j,k) & = & {\hat C}(i,j,k) {\hat C}(i,j,k)^{\dagger} \\
{\hat O}(i,j,{\bar k}) & = & {\hat C}(i,j,{\bar k}) {\hat C}(i,j,
{\bar k})^{\dagger}
\end{eqnarray}
where
\begin{eqnarray}
{\hat C}(i,j,k) & = & \sum_{ijk}{ b_{ij} d_{k} {\hat
a}^{\dagger}_{i} {\hat a}^{\dagger}_{j} {\hat a}^{\dagger}_{k} } \label{eq:23L2} \\
{\hat C}(i,j,{\bar k}) & = & \sum_{ijk}{ b_{ij} d^{*}_{k} {\hat
a}^{\dagger}_{i} {\hat a}^{\dagger}_{j} {\hat a}_{k} } .
\label{eq:23P}
\end{eqnarray}
The notation for the operators ${\hat O}(i,j,k)$ and ${\hat
C}(i,j,k)$ includes their internal summation indices to indicate
succinctly: (i) the ordering of the second-quantized operators with
indices $i$, $j$, and $k$, and (ii) the type of second-quantized
operator with $k$ denoting ${\hat a}^{\dagger}_{k}$ and ${\bar k}$
denoting ${\hat a}_{k}$.  Note that the notation does not indicate
the ordering of the indices on the tensor coefficients which is
alphabetical in both ${\hat C}(i,j,k)$ in Eq.~(\ref{eq:23P}) and
${\hat C}(k,i,j)$ in Eq.~(\ref{eq:23P2}).  Although the summation
indices within ${\hat C}$ and its adjoint are distinct, we only show
primes on the indices of the adjoint when the indices of the two
operators appear in the same sum.  Finally, for the
$N$-representability conditions to be valid for real symmetric and
general Hermitian RDMs, one-index tensors $d_k$ and $d_{\bar k}$
denote $d_{k}$ and $d_{k}^{*}$, respectively. For multi-index
tensors we employ the convention that the first subscript determines
conjugacy, that is $b_{ij..m} = b_{ij..m}$ and $b_{{\bar i}j..m} =
b^{*}_{ij..m}$.

The first operator ${\hat O}(i,j,k)$ arises from lifting the D
condition through the insertion of a {\em particle} projection
operator
\begin{equation}
\sum_{k,k'}{ d_{k} d^{*}_{k'} {\hat a}^{\dagger}_{k} {\hat a}_{k'} }
\end{equation}
while the second operator ${\hat O}(i,j,{\bar k})$ arises from
lifting the D condition through the insertion of a {\em hole}
projection operator
\begin{equation}
\sum_{k,k'}{ d^{*}_{k} d_{k'} {\hat a}_{k} {\hat a}^{\dagger}_{k'} }
.
\end{equation}
The nonnegativity of ${\hat O}(i,j,k)$ and ${\hat O}(i,j,{\bar k})$
generates a pair of {\em lifting conditions} discussed in
Refs.~\cite{M02,HM05}. While these two conditions depend not just on
the 2-RDM but on parts of the 3-RDM, the sum of these two three-body
operators produces a two-body operator
\begin{equation}
\label{eq:L1} ^{2} {\hat O}_{L1} = {\hat O}(i,j,k) + {\hat
O}(i,j,{\bar k}) .
\end{equation}
Because the two-body operator $^{2} {\hat O}_{L1}$ simplifies to the
two-body operator $^{2} {\hat O}_{D}$ in Eq.~(\ref{eq:OD}), its
nonnegativity regenerates the D condition.  With a generalization of
this lifting process, however, we can generate (2,3)-positivity
conditions that are distinct from the known conditions.

We can generalize the lifting process by inserting the creation
operator and the annihilation operator responsible for lifting at
non-adjacent positions.  For example, consider the pair of
three-body operators
\begin{eqnarray}
{\hat O}(k,i,j) & = & {\hat C}(k,i,j) {\hat C}(k,i,j)^{\dagger} \\
{\hat O}({\bar k},i,j) & = & {\hat C}({\bar k},i,j) {\hat C}({\bar
k},i,j)^{\dagger}
\end{eqnarray}
where
\begin{eqnarray}
{\hat C}(k,i,j) & = & \sum_{ijk}{ b_{ij} d_{k} {\hat
a}^{\dagger}_{k}
{\hat a}^{\dagger}_{i} {\hat a}^{\dagger}_{j} } \label{eq:23P2} \\
{\hat C}({\bar k},i,j) & = & \sum_{ijk}{ b_{ij} d^{*}_{k} {\hat
a}_{k} {\hat a}^{\dagger}_{i} {\hat a}^{\dagger}_{j} } .
\end{eqnarray}
In ${\hat O}(k,i,j)$ the creation operator ${\hat a}^{\dagger}_{k}$
in ${\hat C}$ and the annihilation operator ${\hat a}_{k'}$ in the
adjoint of ${\hat C}$, which perform the lifting of the D condition,
are separated from each other by four second-quantized operators;
similarly, in ${\hat O}({\bar k},i,j)$ the creation and annihilation
operators, ${\hat a}_{k}$ and ${\hat a}^{\dagger}_{k'}$
respectively, are separated from each other by four second-quantized
operators. Because the components of the projectors are separated,
the nonnegativity of ${\hat O}(k,i,j)$ and ${\hat O}({\bar k},i,j)$
generates a pair of generalized lifting conditions that extend those
discussed in Refs.~\cite{M02,HM05}.

While individually ${\hat O}(k,i,j)$ and ${\hat O}({\bar k},i,j)$
depend on three-particle operators, their sum generates a two-body
operator
\begin{equation}
\label{eq:L2} ^{2} {\hat O}_{L2} = {\hat O}(k,i,j) + {\hat O}({\bar
k},i,j) .
\end{equation}
Unlike $^{2} {\hat O}_{L1}$, the nonnegativity of the lifted
operator $^{2} {\hat O}_{L2}$ is not necessarily implied by the D,
Q, G, T1, and T2 conditions.  Importantly, $^{2} {\hat O}_{L2}$ does
not simply rearrange to $^{2} {\hat O}_{L1}$ because the creation
and annihilation operators are non-commutative.  Based on the
possible orderings of the fundamental second-quantized operators,
there are nine distinct ways to lift the D condition while canceling
the resulting three-particle operators and hence, nine distinct
lifting conditions from the D condition.  Similarly, there are nine
distinct (2,3)-positivity conditions from lifting the Q condition
and nine from lifting the G condition.  Three of these 27 lifting
conditions reduce to the D, Q, and G conditions, respectively while
the other conditions are distinct because the second-quantized
operators in quantum mechanics form a non-commutative algebra.

Table~I summarizes the (2,3)-positivity conditions by giving a
representative condition from each of the two classes: (i) the
lifting conditions and (ii) the pure conditions.  While the lifting
conditions arise from lifting the 2-positivity conditions, the pure
conditions cannot be obtained from lifting any of the lower
conditions.  Table~I gives nonnegativity of $^{2} {\hat O}_{L2}$ and
the T1 condition as representative conditions of the lifting and
pure (2,3)-positivity conditions, respectively.  All of the other
(2,3)-conditions can be obtained from these representative
conditions through two processes, (i) {\em switching} of the
second-quantized operators in the ${\hat C}(i,j,k)$ between creators
and annihilators and (ii) {\em reordering} of the second-quantized
operators in the ${\hat C}(i,j,k)$.


\begin{table*}[ht!]

\caption{The (2,4)-positivity conditions can be derived from conic
(linear nonnegative) combinations of the (4,4)-positivity conditions
that cancel the 3- and 4-particle operators.}

\label{t:24}

\begin{ruledtabular}
\begin{tabular}{cccc}

Class & Type & Representative Condition & ${\hat C}$ Definition \\

\hline

1 & Lifted (2,2) & ${\rm Tr}( ({\hat O}(l,k,i,j) + {\hat O}(l,{\bar
k},i,j) + {\hat O}({\bar l},k,i,j) + {\hat O}({\bar l},{\bar k},i,j)
) \, {}^{2} D)
\ge 0$ & Eq.~(\ref{eq:24L2}) \\

2 & Lifted (2,3) & ${\rm Tr}( ({\hat O}(l,i,j,k) + {\hat O}(l,{\bar
i},{\bar j},{\bar k}) + {\hat O}({\bar l},{\bar i},{\bar j},{\bar
k}) + {\hat O}({\bar l},i,j,k) ) \, {}^{2} D)
\ge 0$ & Eq.~(\ref{eq:24L3}) \\

3  & Pure (2,4) & ${\rm Tr}( (3 {\hat O}(i,j,k,l) + {\hat
O}(i,j,k,{\bar l}) + {\hat O}(i,j,{\bar k},l) + {\hat O}(i,{\bar
j},k,l) + {\hat O}({\bar i},j,k,l) + {\hat O}({\bar i},{\bar
j},{\bar k},{\bar l})) \, {}^{2} D) \ge 0$ & Eq.~(\ref{eq:24P}) \\

\end{tabular}
\end{ruledtabular}

\end{table*}

Switching the second-quantized operators with index $j$ in the L2
condition of Eq.~(\ref{eq:L2}), for example, generates a lifted G
condition
\begin{equation}
 ^{2} {\hat O}_{L3} = {\hat O}(k,i,{\bar j}) + {\hat O}({\bar
k},i,{\bar j}) .
\end{equation}
Note that switching the second-quantized operators with index $k$ in
Eq.~(\ref{eq:L2}) simply regenerates the same condition while
switching the second-quantized operators associated with indices $i$
and $j$ generates lifted G and Q conditions from the lifted D
condition.  Reordering of the second-quantized operators in the L2
condition of Eq.~(\ref{eq:L2}), by contrast, produces the other 9
lifted D conditions; for example, reordering L2 yields the L1
condition in Eq.~(\ref{eq:L1}).  Similarly, for the pure
(2,3)-positivity conditions switching of the second-quantized
operators with index $k$ in T1 produces the T2 condition in
Eq.~(\ref{eq:T2}).  The other two distinct T2 conditions, ${\bar
T2}$ and ${\tilde T2}$, in Eqs.~(\ref{eq:T2b}) and~(\ref{eq:T2t})
are generated not by switching but by reordering the
second-quantized operators in the T2 condition of Eq.~(\ref{eq:T2}).

\subsubsection{(2,4)-positivity conditions}


The (2,4)-positivity conditions, arising from considering all ${\hat
C}_{i}$ operators of degree less than or equal to four in
Eq.~(\ref{eq:O2}), consist of two classes of lifting conditions and
one class of pure conditions, which are summarized in Table~II.  The
two classes of lifting conditions are generated from lifting the two
classes of (2,3)-positivity conditions.  As in the previous section,
the generalized lifting is performed by (i) inserting a creation
operator into each ${\hat C}_{i}$ operator contributing to the
condition, (ii) converting the inserted creation operator into an
annihilation operator in the operator produced from step (i), and
(iii) adding the two lifted operators from steps (i) and (ii)
together to produce a two-particle operator.  The nonnegativity of
the resulting two-particle operator generates a lifting
(2,4)-positivity condition.  Representative lifting conditions for
both classes are shown in Table~II.  The ${\hat C}$ operators in the
first and second classes of lifted (2,3)-positivity conditions and
the pure (2,4)-positivity condition are given by
\begin{eqnarray}
{\hat C}(l,k,i,j) & = & \sum_{ijkl}{ b_{ij} d_{k} e_{l}
{\hat a}^{\dagger}_{l} {\hat a}^{\dagger}_{k} {\hat a}^{\dagger}_{i} {\hat a}^{\dagger}_{j} } \label{eq:24L2} \\
{\hat C}(l,i,j,k) & = & \sum_{ijkl}{ b_{ijk} d_{l} {\hat
a}^{\dagger}_{l}
{\hat a}^{\dagger}_{i} {\hat a}^{\dagger}_{j} {\hat a}^{\dagger}_{k} }  \label{eq:24L3} \\
{\hat C}(i,j,k,l) & = & \sum_{ijkl}{ b_{i} d_{j} e_{k} f_{l} {\hat
a}^{\dagger}_{i} {\hat a}^{\dagger}_{j} {\hat a}^{\dagger}_{k} {\hat
a}^{\dagger}_{l} } \label{eq:24P}
\end{eqnarray}
where $b_{i}$, $d_{i}$, $e_{i}$, $f_{i}$, $b_{ij}$, $b_{ijk}$, and
${\hat a}^{\dagger}_{i}$ become $b_{i}^{*}$, $d_{i}^{*}$,
$e_{i}^{*}$, $f_{i}^{*}$, $b_{ij}^{*}$, $b_{ijk}^{*}$, and ${\hat
a}_{i}$ when $i={\bar i}$.  The rank of the largest tensor changes
from three in Eq.~(\ref{eq:24L2}) to one in Eq.~(\ref{eq:24P}) to
effect the cancelation of the 3- and 4-RDOs in the combinations of
operators in Table~II.  Fusing the tensors $b_{i}$ and $d_{j}$ into
a single rank-two tensor $b_{ij}$ in Eq.~(\ref{eq:24P}), for
example, would cause the operator combinations in Table~II to depend
on the 3- and 4-RDOs.  Additional (2,4)-positivity conditions can be
generated from the representative conditions through a combination
of switching and reordering of the creation and annihilation
operators.

The pure (2,4)-positivity conditions, presented in Ref.~\cite{M12b},
depend upon only the 2-RDM through conic combinations that cancel
the 3- and 4-RDMs. As in the (2,3)-conditions, the cancelations
depend upon the conic combination of pairs of operators that differ
from each other by an odd number of switchings---exchanges of
creators and annihilators. Generating an extreme condition on the
2-RDM requires that we consider the minimum number of conic
combinations that effect the cancelation of the higher RDMs.  Each
pure (2,4)-positivity condition involves the conic combination of
eight four-particle operators by Eq.~(\ref{eq:O2}).  These eight
four-particle operators can be grouped into the four pairs that
depend upon only three-particle operators:
\begin{eqnarray}
{\hat O}({\bar i},j,k,l) & + & {\hat O}({\bar i},{\bar j},{\bar k},
{\bar l}) \\
{\hat O}(i,{\bar j},k,l) & + & {\hat O}(i,j,k,l) \\
{\hat O}(i,j,{\bar k},l) & + & {\hat O}(i,j,k,l) \\
{\hat O}(i,j,k,{\bar l}) & + & {\hat O}(i,j,k,l)
\end{eqnarray}
The operators in the first pair differ from each other by the
switching of three creation and annihilation operators while the
operators in the other three pairs differ from each other by the
switching of one creation operator and one annihilation operator.
Rearranging the second-quantized operators in the four pairings into
normal order with creators to the left of the annihilators generates
expressions involving the sum of 9, 5, 3, and 1 3-RDOs,
respectively.  Upon summation the 9 3-RDOs from the one pairing with
three switchings cancels with the 5, 3, and 1 3-RDOs from the three
pairings with one switching, and hence the final operator depends
upon only the 2-RDO.


\begin{table*}[t!]

\caption{The representative pure (2,4)-positivity condition $g_{1}
\ge 0$ as well as three other conditions generated from its
reordering, $g_{2} \ge 0$, $g_{3} \ge 0$, and $g_{4} \ge 0$, are
presented.  Unlike the situation in the classical limit, in the
quantum case additional conditions can be generated from each of the
16~conditions obtained from switching by reordering the creation and
annihilation operators while preserving the cancelation of the 3-
and 4-particle operators.}

\label{t:24r}

\begin{ruledtabular}
\begin{tabular}{cc}

 Condition & ${\hat C}$ Definition \\

\hline

$g_{1}({}^{2} D) = {\rm Tr}( (3 {\hat O}(i,j,k,l) + {\hat
O}(i,j,k,{\bar l}) + {\hat O}(i,j,{\bar k},l) + {\hat O}(i,{\bar
j},k,l) + {\hat O}({\bar i},j,k,l) + {\hat O}({\bar i},{\bar
j},{\bar k},{\bar l})) \, {}^{2} D)
\ge 0$ & Eq.~(\ref{eq:24P}) \\

$g_{2}({}^{2} D) = {\rm Tr}( (3 {\hat O}(i,j,k,l) + {\hat
O}(i,j,k,{\bar l}) + {\hat O}(i,{\bar k},j,l) + {\hat O}(k,{\bar
j},i,l) + {\hat O}(j,{\bar i},k,l) + {\hat O}({\bar i},{\bar
j},{\bar k},{\bar l})) \, {}^{2} D) \ge 0$ & Eq.~(\ref{eq:24P}) \\

$g_{3}({}^{2} D) = {\rm Tr}( (3 {\hat O}(i,j,k,l) + {\hat
O}(i,j,{\bar l},k) + {\hat O}(i,l,{\bar k},j) + {\hat O}(k,{\bar
j},i,l) + {\hat O}(j,{\bar i},k,l) + {\hat O}({\bar i},{\bar
j},{\bar k},{\bar l})) \, {}^{2} D) \ge 0$ & Eq.~(\ref{eq:24P}) \\

$g_{4}({}^{2} D) = {\rm Tr}( (3 {\hat O}(i,j,k,l) + {\hat
O}(i,j,k,{\bar l}) + {\hat O}(i,{\bar k},j,l) + {\hat O}(k,{\bar
j},i,l) + {\hat O}(j,{\bar i},k,l) + {\hat O}({\bar i},{\bar
j},{\bar k},{\bar l})) \, {}^{2} D) \ge 0$ & Eq.~(\ref{eq:24P})

\end{tabular}
\end{ruledtabular}

\end{table*}

Other pure (2,4)-positivity conditions can be generated from the
representative condition through switching and reordering of the
second-quantized operators.  To maintain the cancelation of the 3-
and 4-RDOs, we must perform the same switching of creation and
annihilation operators in each operator ${\hat C}(i,j,k,l)$
contributing to the condition.  Because each fundamental
second-quantized operator can be either a creation or an
annihilation operator, there are $2^{4}$ or 16 conditions from
switching.  Eight of these conditions can be generated from the
other 8 conditions by switching all creation and annihilation
operators by particle-hole symmetry.  In the limit that the
expansion coefficients $b_{i}$, $d_{j}$, $e_{k}$, and $f_{l}$ become
orthogonal unit vectors, these 16 conditions reduce to the 16
conditions in (2,4)-class of the classical (or diagonal)
$N$-representability problem~\cite{MD72,E78,Cuts}.  The quantum
mechanical formulation of these conditions, however, is much more
general because the expansion coefficients need not be orthogonal.
When the expansion coefficients are non-orthogonal, the creation and
annihilation operators become non-commutative operators, and hence,
the conditions depend upon their ordering.

In the quantum case additional conditions can be generated from each
of the 16 conditions by reordering the creation and annihilation
operators while preserving the cancelation of the 3- and 4-RDOs.
These additional conditions are related to the original 16
conditions as the generalized T2 conditions are related to the T2
condition in the (2,3)-positivity conditions.  Table~III presents
the representative pure (2,4)-positivity condition as well as three
other conditions generated from its reordering.  Each of these four
conditions differs from the others by a few terms involving the
2-RDM.  For example, the first and second conditions differ by only
one term
\begin{equation}
g_{2} = g_{1} + 4 \Re \left ( \alpha \beta \sum_{ik;i'l'}{ b_{i}
e_{k} \, {}^{2} D^{ik}_{i'l'} \, b^{*}_{i'} f^{*}_{l'} } \right )
\ge 0.
\end{equation}
where
\begin{eqnarray}
\alpha & = & \sum_{j}{d_{j} e^{*}_{j}} \\
\beta  & = & \sum_{j}{f_{j} d^{*}_{j}}
\end{eqnarray}
and $\Re$ selects the real part of the expression.  When this term
is negative, inequality $g_{2}$ is stronger than $g_{1}$, but when
this term is positive, inequality $g_{1}$ is stronger than $g_{2}$.
In the classical case, where the expansion coefficients are
orthogonal, these two conditions are equivalent because both
$\alpha$ and $\beta$ are zero, and hence, this additional term
vanishes.

\subsubsection{(2,5)-positivity conditions}



\begin{table*}[t!]

\caption{The (2,5)-positivity conditions can be derived from conic
(linear nonnegative) combinations of the (5,5)-positivity conditions
that cancel the 3, 4-, and 5-particle operators.}

\label{t:25}

\begin{ruledtabular}
\begin{tabular}{cccc}

Class & Type & Representative Condition & ${\hat C}$ Definition \\

\hline

1 & Lifted (2,2) & ${\rm Tr}( ({\hat O}(m,l,k,i,j) + {\hat
O}(m,{\bar l},k,i,j) + {\hat O}({\bar m},l,k,i,j) + {\hat
O}({\bar m},{\bar l},k,i,j)$ & Eq.~(\ref{eq:25L2}) \\

  &              & $ ~~~~~~~~~~~~~~ + {\hat O}(m,l,{\bar k},i,j) +
  {\hat O}(m,{\bar l},{\bar k},i,j) + {\hat O}({\bar m},l,{\bar
  k},i,j) + {\hat O}({\bar m},{\bar l},{\bar k},i,j) ) \, {}^{2} D)
\ge 0$ &  \\

2 & Lifted (2,3) & ${\rm Tr}( ({\hat O}(m,l,i,j,k) + {\hat
O}(m,l,{\bar i},{\bar j},{\bar k}) + {\hat O}({\bar m},l,{\bar
i},{\bar j},{\bar k}) + {\hat O}({\bar m},l,i,j,k)$ &
Eq.~(\ref{eq:25L3}) \\

  &              & $~~~~~~~~~~~~~~ + {\hat O}(m,{\bar l},i,j,k) +
{\hat O}(m,{\bar l},{\bar i},{\bar j},{\bar k}) + {\hat O}({\bar
m},{\bar l},{\bar i},{\bar j},{\bar k}) + {\hat O}({\bar m},{\bar
l},i,j,k) ) \, {}^{2} D)
\ge 0$ &  \\

3 & Lifted (2,4) & ${\rm Tr}( (3 {\hat O}(m,i,j,k,l) + {\hat
O}(m,i,j,k,{\bar l}) + {\hat O}(m,i,j,{\bar k},l)$ & Eq.~(\ref{eq:25L4}) \\

  &              & $~~~ + {\hat O}(m,i,{\bar j},k,l) + {\hat O}(m,{\bar i},j,k,l)
+ {\hat O}(m,{\bar i},{\bar j},{\bar k},{\bar l})$ &  \\

  &              & $~+ 3 {\hat O}({\bar m},i,j,k,l) + {\hat
O}({\bar m},i,j,k,{\bar l}) + {\hat O}({\bar m},i,j,{\bar k},l)$ &   \\

  &              & $~~~~~~~~~~~~~~~+ {\hat O}({\bar m},i,{\bar j},k,l) +
  {\hat O}({\bar m},{\bar i},j,k,l) + {\hat O}({\bar m},{\bar i},{\bar j},{\bar k},{\bar l}))
  \, {}^{2} D) \ge 0$ &  \\

4 & Pure (2,5)   & ${\rm Tr}( (3 {\hat O}(i,j,k,l,m) + {\hat
O}(i,j,k,l,{\bar m})
  + {\hat O}(i,j,k,{\bar l},m)$ & Eq.~(\ref{eq:25P}) \\

  &              & $~~~ + {\hat O}(i,j,{\bar k},l,m) + {\hat O}(i,{\bar j},k,l,m)
+ {\hat O}({\bar i},j,k,l,m)$ &  \\

  &              & $~+ 3 {\hat O}({\bar i},{\bar j},{\bar k},{\bar l},{\bar m}) +
  {\hat O}({\bar i},{\bar j},{\bar k},{\bar l},m) + {\hat O}({\bar i},{\bar j},{\bar k},
l,{\bar m})$ &  \\

  &              & $~~~~~~~~~~~~~~~ + {\hat O}({\bar i},{\bar j},k,{\bar l},{\bar m}) +
  {\hat O}({\bar i},j,{\bar k},{\bar l},{\bar m}) + {\hat O}(i,{\bar j},{\bar k},{\bar l},{\bar m}))
  \, {}^{2} D) \ge 0$ &  \\

5 & Pure (2,5)   & ${\rm Tr}( (6 {\hat O}(i,j,k,l,m) + 3 {\hat
O}(i,j,k,l,{\bar m}) + 3 {\hat O}(i,j,k,{\bar l},m)$ & Eq.~(\ref{eq:25P}) \\

  &              & $~~ + 3 {\hat O}(i,j,{\bar k},l,m) + 3 {\hat O}(i,{\bar j},k,l,m)
+ 3 {\hat O}({\bar i},j,k,l,m)$ &  \\

  &              & $~+ {\hat O}(i,j,k,{\bar l},{\bar m}) +
  {\hat O}(i,j,{\bar k},l,{\bar m}) + {\hat O}(i,j,{\bar k},
{\bar l},m)$ &  \\

  &              & $~+ {\hat O}(i,{\bar j},k,l,{\bar m}) +
  {\hat O}(i,{\bar j},k,{\bar l},m) + {\hat O}(i,{\bar j},{\bar k},
l,m)$ &  \\

  &              & $~+ {\hat O}({\bar i},j,k,l,{\bar m}) +
  {\hat O}({\bar i},j,k,{\bar l},m) + {\hat O}({\bar i},j,{\bar k},
l,m)$ &  \\

  &              & $ + {\hat O}({\bar i},{\bar j},k,l,m) +
{\hat O}({\bar i},{\bar j},{\bar k},{\bar l},{\bar m})) \, {}^{2} D)
\ge 0$ &  \\

6 & Pure (2,5)   & ${\rm Tr}( (6 {\hat O}(i,j,k,l,m) + 3 {\hat
O}(i,j,k,l,{\bar m}) + 3 {\hat O}(i,j,k,{\bar l},m)$ & Eq.~(\ref{eq:25P}) \\

  &              & $~~ + 3 {\hat O}(i,j,{\bar k},l,m) + 3 {\hat O}(i,{\bar j},k,l,m)
+ {\hat O}({\bar i},j,k,l,m)$ &  \\

  &              & $~+ {\hat O}(i,j,k,{\bar l},{\bar m}) +
  {\hat O}(i,j,{\bar k},l,{\bar m}) + {\hat O}(i,j,{\bar k},
{\bar l},m)$ &  \\

  &              & $~+ {\hat O}(i,{\bar j},k,l,{\bar m}) +
  {\hat O}(i,{\bar j},k,{\bar l},m) + {\hat O}(i,{\bar j},{\bar k},
l,m)$ &  \\

  &              & $~+ {\hat O}(i,j,k,l,{\bar m}) +
  {\hat O}(i,j,k,{\bar l},m) + {\hat O}(i,j,{\bar k},
l,m)$ &  \\

  &              & $ + {\hat O}(i,{\bar j},k,l,m) +
3 {\hat O}({\bar i},{\bar j},{\bar k},{\bar l},{\bar m})) \, {}^{2}
D) \ge 0$ &

\end{tabular}
\end{ruledtabular}

\end{table*}

The (2,5)-positivity conditions are generated from considering all
${\hat C}_{i}$ operators of degree less than or equal to five in
Eq.~(\ref{eq:O2}). These conditions consist of three classes of
lifting conditions and three classes of pure conditions, which are
given in Table~IV.  The lifting conditions arise from lifting the
three different classes of (2,4)-positivity conditions.  The ${\hat
C}$ operators of the first, second, and third classes of lifting
conditions are given by
\begin{eqnarray}
{\hat C}(m,l,k,i,j) & = & \sum_{ijklm}{ b_{ij} d_{k} e_{l} f_{m}
{\hat a}^{\dagger}_{m} {\hat a}^{\dagger}_{l} {\hat a}^{\dagger}_{k}
{\hat a}^{\dagger}_{i}
{\hat a}^{\dagger}_{j} } \label{eq:25L2} \\
{\hat C}(m,l,i,j,k) & = & \sum_{ijklm}{ b_{ijk} d_{l} e_{m} {\hat
a}^{\dagger}_{m} {\hat a}^{\dagger}_{l} {\hat a}^{\dagger}_{i} {\hat
a}^{\dagger}_{j} {\hat a}^{\dagger}_{k} } \label{eq:25L3} \\
{\hat C}(m,i,j,k,l) & = & \sum_{ijklm}{ b_{i} d_{j} e_{k} f_{l}
g_{m} {\hat a}^{\dagger}_{m} {\hat a}^{\dagger}_{i} {\hat
a}^{\dagger}_{j} {\hat a}^{\dagger}_{k} {\hat a}^{\dagger}_{l} },
\label{eq:25L4}
\end{eqnarray}
respectively, and the ${\hat C}$ operators of the three classes of
pure conditions are given by
\begin{equation}
{\hat C}(i,j,k,l,m) = \sum_{ijklm}{ b_{i} d_{j} e_{k} f_{l} g_{m}
{\hat a}^{\dagger}_{i} {\hat a}^{\dagger}_{j} {\hat a}^{\dagger}_{k}
{\hat a}^{\dagger}_{l} {\hat a}^{\dagger}_{m} } \label{eq:25P}
\end{equation}
where the $b_{i}$, $d_{i}$, $e_{i}$, $f_{i}$, $g_{i}$, $b_{ij}$,
$b_{ijk}$, and ${\hat a}^{\dagger}_{i}$ become $b_{i}^{*}$,
$d_{i}^{*}$, $e_{i}^{*}$, $f_{i}^{*}$, $g_{i}^{*}$, $b_{ij}^{*}$,
$b_{ijk}^{*}$, and ${\hat a}_{i}$ when $i={\bar i}$.  Note that the
operators in Eqs.~(\ref{eq:25L4}) and~(\ref{eq:25P}) are not
equivalent after switching.  Switching of creators to annihilators
in the ${\hat C}$ operators in the representative conditions in
Table~IV produces 16, 32, and 32 conditions in the pure classes 4,
5, and 6, respectively. Class~4 has fewer conditions because its
conditions, unlike those in classes~5 and~6, possess particle-hole
symmetry. Particle-hole symmetry is present in all of the pure
(2,3)-positivity conditions and none of the pure (2,4)-conditions.
Additional conditions can be generated from the representative
conditions through reordering of the creation and annihilation
operators.  Like the (2,3)- and (2,4)-positivity conditions, the
(2,5)-conditions generate all of the classical (diagonal)
$N$-representability conditions when the expansion coefficients
$b_{i}$, $d_{j}$, $e_{k}$, $f_{l}$, and $g_{m}$ are chosen to be
orthogonal unit vectors.

\subsubsection{(2,6)-positivity conditions}



\begin{table*}[t!]

\caption{The (2,6)-positivity conditions can be derived from conic
combinations of the (6,6)-positivity conditions that cancel the 3,
4-, 5-, and 6-particle operators.  There are 6 classes of lifting
conditions (not shown) and 18 classes of pure conditions (shown).
The table provides a representative operator for each of the 18
classes.  The conic combination of all 32 pairs with the weights in
the $x^{\rm th}$ column generates a representative operator for
class~$x$.  The trace of each representative operator against the
2-RDM generates a representative condition on the 2-RDM.}

\label{t:26}

\begin{ruledtabular}
\begin{tabular}{ccccccccccccccccccc}

          & \multicolumn{18}{c}{Weights ($\alpha/\beta$)} \\
          \cline{2-19}

Operators & 1 & 2 & 3 & 4 & 5 & 6 & 7 & 8 & 9 & 10 & 11 & 12 & 13 &
14 & 15 & 16 & 17 & 18\\

\hline

$\alpha {\hat O}(ijklmn)+ \beta {\hat O}({\bar i}{\bar j}{\bar
k}{\bar l}{\bar m}{\bar n})$ & 20/2 & 12/6 & 20/6 & 12/12 & 20/12 &
20/12 & 9/6 & 5/3 & 9/5 & 9/3 & 5/2 & 14/9 & 14/6 & 14/3 &
6/3 & 6/5 & 1/6 & 3/3 \\
$\alpha {\hat O}(ijklm{\bar n})+ \beta {\hat O}({\bar i}{\bar
j}{\bar k}{\bar l}{\bar m}n)$ & 12/0 & 6/2 & 12/2 & 6/6 & 12/6 &
12/6 & 5/3 & 2/1 & 3/1 & 5/1 & 3/1 & 9/5 & 9/3 & 9/1 & 5/2 &
5/3 & 0/3 & 2/1 \\
$\alpha {\hat O}(ijkl{\bar m}n)+ \beta {\hat O}({\bar i}{\bar
j}{\bar k}{\bar l}m{\bar n})$ & 12/0 & 6/2 & 12/2 & 6/6 & 12/6 &
12/6 & 6/3 & 3/1 & 6/3 & 6/1 & 3/1 & 9/5 & 9/3 & 9/1 & 3/1 & 3/3 &
0/3 & 1/1 \\
$\alpha {\hat O}(ijkl{\bar m}{\bar n})+ \beta {\hat O}({\bar i}{\bar
j}{\bar k}{\bar l}mn)$ & 6/0 & 2/0 & 6/0 & 2/2 & 6/2 & 6/2 & 3/1 &
1/0 & 1/0 & 3/0 & 1/0 & 5/2 & 5/1 &
5/0 & 3/1 & 2/1 & 0/1 & 1/0 \\
$\alpha {\hat O}(ijk{\bar l}mn)+ \beta {\hat O}({\bar i}{\bar
j}{\bar k}l{\bar m}{\bar n})$ & 12/0 & 6/2 & 12/2 & 6/6 & 12/6 &
12/6 & 5/3 & 2/1 & 5/2 & 5/1 & 3/1 & 9/5 & 9/3 & 9/1 & 3/1 &
3/2 & 0/3 & 3/2 \\
$\alpha {\hat O}(ijk{\bar l}m{\bar n})+ \beta {\hat O}({\bar i}{\bar
j}{\bar k}l{\bar m}n)$ & 6/0 & 2/0 & 6/0 & 2/2 & 6/2 & 6/2 & 2/1 &
0/0 & 1/0 & 2/0 & 1/0 & 5/2 & 5/1 & 5/0 & 2/0 & 3/1 &
0/1 & 3/1 \\
$\alpha {\hat O}(ijk{\bar l}{\bar m}n)+ \beta {\hat O}({\bar i}{\bar
j}{\bar k}lm{\bar n})$ & 6/0 & 2/0 & 6/0 & 2/2 & 6/2 & 6/2 & 3/1 &
1/0 & 3/1 & 3/0 & 2/1 & 5/2 & 5/1 & 5/0 & 1/0 & 1/1 & 0/1 & 2/1 \\
$\alpha {\hat O}(ijk{\bar l}{\bar m}{\bar n})+ \beta {\hat O}({\bar
i}{\bar j}{\bar k}lmn)$ & 2/2 & 0/0 & 2/0 & 0/0 & 2/0 & 2/0 & 1/0 &
0/0 & 0/0 & 1/0 & 0/0 & 2/0 & 2/0 & 2/0 & 1/0 & 1/0 & 1/0 & 3/1 \\
$\alpha {\hat O}(ij{\bar k}lmn)+ \beta {\hat O}({\bar i}{\bar
j}k{\bar l}{\bar m}{\bar n})$ & 12/0 & 6/2 & 12/2 & 6/6 & 12/6 &
12/6 & 3/2 & 3/2 & 6/3 & 6/2 & 3/1 & 6/3 & 9/3 & 6/0 & 3/1 & 3/2
& 1/5 & 2/3 \\
$\alpha {\hat O}(ij{\bar k}lm{\bar n})+ \beta {\hat O}({\bar i}{\bar
j}k{\bar l}{\bar m}n)$ & 6/0 & 2/0 & 6/0 & 2/2 & 6/2 & 6/2 & 1/1 &
1/1 & 2/1 & 3/1 & 2/1 & 3/1 & 5/1 & 3/0 & 3/1 & 3/1 & 0/2 & 1/1 \\
$\alpha {\hat O}(ij{\bar k}l{\bar m}n)+ \beta {\hat O}({\bar i}{\bar
j}k{\bar l}m{\bar n})$ & 6/0 & 2/0 & 6/0 & 2/2 & 6/2 & 6/2 & 1/0 &
1/0 & 3/1 & 3/0 & 1/0 & 3/1 & 5/1 & 3/0 & 1/0 & 1/1 & 0/2 & 0/1 \\
$\alpha {\hat O}(ij{\bar k}l{\bar m}{\bar n})+ \beta {\hat O}({\bar
i}{\bar j}k{\bar l}mn)$ & 2/2 & 0/0 & 2/0 & 0/0 & 2/0 & 2/0 & 0/0 &
0/0 & 0/0 & 1/0 & 0/0 & 1/0 & 2/0 & 1/1 & 2/1 & 1/0 &
0/0 & 0/0 \\
$\alpha {\hat O}(ij{\bar k}{\bar l}mn)+ \beta {\hat O}({\bar i}{\bar
j}kl{\bar m}{\bar n})$ & 6/0 & 2/0 & 6/0 & 2/2 & 6/2 & 6/2 & 1/1 &
1/1 & 3/1 & 3/1 & 2/1 & 3/1 & 5/1 & 3/0 & 1/0 & 1/0 &
1/3 & 1/1 \\
$\alpha {\hat O}(ij{\bar k}{\bar l}m{\bar n})+ \beta {\hat O}({\bar
i}{\bar j}kl{\bar m}n)$ & 2/2 & 0/0 & 2/0 & 0/0 & 2/0 & 2/0 & 0/1 &
0/1 & 1/1 & 1/1 & 1/1 & 1/0 & 2/0 & 1/1 & 1/0 & 2/0 &
1/1 & 1/0 \\
$\alpha {\hat O}(ij{\bar k}{\bar l}{\bar m}n)+ \beta {\hat O}({\bar
i}{\bar j}klm{\bar n})$ & 2/2 & 0/0 & 2/0 & 0/0 & 2/0 & 2/0 & 0/0 &
0/0 & 1/0 & 1/0 & 1/1 & 1/0 & 2/0 & 1/1 & 0/0 & 0/0 &
1/1 & 0/0 \\
$\alpha {\hat O}(i{\bar j}klmn)+ \beta {\hat O}({\bar i}j{\bar
k}{\bar l}{\bar m}{\bar n})$ & 12/0 & 6/2 & 12/2 & 6/6 & 6/2 & 12/6
& 3/1 & 3/1 & 5/2 & 5/1 & 2/0 & 3/1 & 3/0 & 9/1 & 1/0 & 1/1 &
1/5 & 1/1 \\
$\alpha {\hat O}(i{\bar j}klm{\bar n})+ \beta {\hat O}({\bar
i}j{\bar k}{\bar l}{\bar m}n)$ & 6/0 & 2/0 & 6/0 & 2/2 & 2/0 & 6/2 &
1/0 & 1/0 & 1/0 & 2/0 & 1/0 & 1/0 & 1/0 & 5/0 & 1/0 & 1/0 &
1/3 & 1/0 \\
$\alpha {\hat O}(i{\bar j}kl{\bar m}n)+ \beta {\hat O}({\bar
i}j{\bar k}{\bar l}m{\bar n})$ & 6/0 & 2/0 & 6/0 & 2/2 & 2/0 & 6/2 &
2/0 & 2/0 & 3/1 & 3/0 & 1/0 & 1/0 & 1/0 & 5/0 & 0/0 & 0/1 &
1/3 & 0/0 \\
$\alpha {\hat O}(i{\bar j}kl{\bar m}{\bar n})+ \beta {\hat O}({\bar
i}j{\bar k}{\bar l}mn)$ & 2/2 & 0/0 & 2/0 & 0/0 & 0/0 & 2/0 & 1/0 &
1/0 & 0/0 & 1/0 & 0/0 & 0/0 & 0/1 & 2/0 & 1/1 & 0/0 &
2/2 & 1/0 \\
$\alpha {\hat O}(i{\bar j}k{\bar l}mn)+ \beta {\hat O}({\bar
i}j{\bar k}l{\bar m}{\bar n})$ & 6/0 & 2/0 & 6/0 & 2/2 & 2/0 & 6/2 &
1/0 & 1/0 & 2/0 & 2/0 & 1/0 & 1/0 & 1/0 & 5/0 & 0/0 & 0/0 &
0/2 & 1/0 \\
$\alpha {\hat O}(i{\bar j}k{\bar l}m{\bar n})+ \beta {\hat O}({\bar
i}j{\bar k}l{\bar m}n)$ & 2/2 & 0/0 & 2/0 & 0/0 & 0/0 & 2/0 & 0/0 &
0/0 & 0/0 & 0/0 & 0/0 & 0/0 & 0/1 & 2/0 & 0/0 & 1/0 &
1/1 & 2/0 \\
$\alpha {\hat O}(i{\bar j}k{\bar l}{\bar m}n)+ \beta {\hat O}({\bar
i}j{\bar k}lm{\bar n})$ & 2/2 & 0/0 & 2/0 & 0/0 & 0/0 & 2/0 & 1/0 &
1/0 & 1/0 & 1/0 & 1/1 & 0/0 & 0/1 & 2/0 & 0/1 & 0/1 &
1/1 & 1/0 \\
$\alpha {\hat O}(i{\bar j}{\bar k}lmn)+ \beta {\hat O}({\bar
i}jk{\bar l}{\bar m}{\bar n})$ & 6/0 & 2/0 & 6/0 & 2/2 & 2/0 & 6/2 &
0/0 & 2/1 & 3/1 & 3/1 & 1/0 & 0/0 & 1/0 & 3/0 & 0/0 & 0/0 &
0/3 & 1/2 \\
$\alpha {\hat O}(i{\bar j}{\bar k}lm{\bar n})+ \beta {\hat O}({\bar
i}jk{\bar l}{\bar m}n)$ & 2/2 & 0/0 & 2/0 & 0/0 & 0/0 & 2/0 & 0/1 &
1/1 & 1/1 & 1/1 & 1/1 & 0/1 & 0/1 & 1/1 & 1/1 & 1/0 &
0/1 & 1/1 \\
$\alpha {\hat O}(i{\bar j}{\bar k}l{\bar m}n)+ \beta {\hat O}({\bar
i}jk{\bar l}m{\bar n})$ & 2/2 & 0/0 & 2/0 & 0/0 & 0/0 & 2/0 & 0/0 &
1/0 & 1/0 & 1/0 & 0/0 & 0/1 & 0/1 & 1/1 & 0/1 & 0/1 &
0/1 & 0/1 \\
$\alpha {\hat O}(i{\bar j}{\bar k}{\bar l}mn)+ \beta {\hat O}({\bar
i}jkl{\bar m}{\bar n})$ & 2/2 & 0/0 & 2/0 & 0/0 & 0/0 & 2/0 & 0/1 &
1/1 & 1/0 & 1/1 & 1/1 & 0/1 & 0/1 & 1/1 & 0/1 & 0/0 &
0/1 & 0/0 \\
$\alpha {\hat O}({\bar i}jklmn)+ \beta {\hat O}(i{\bar j}{\bar
k}{\bar l}{\bar m}{\bar n})$ & 12/0 & 6/2 & 6/0 & 2/2 & 6/2 & 2/0 &
5/3 & 3/2 & 3/1 & 3/0 & 3/1 & 6/3 & 6/1 & 6/0 & 5/3 & 3/3
& 0/3 & 1/2 \\
$\alpha {\hat O}({\bar i}jklm{\bar n})+ \beta {\hat O}(i{\bar
j}{\bar k}{\bar l}{\bar m}n)$ & 6/0 & 2/0 & 2/0 & 0/0 & 2/0 & 0/0 &
2/1 & 1/1 & 0/0 & 1/0 & 2/1 & 3/1 & 3/0 & 3/0 & 3/1 & 2/1 &
0/1 & 0/0 \\
$\alpha {\hat O}({\bar i}jkl{\bar m}n)+ \beta {\hat O}(i{\bar
j}{\bar k}{\bar l}m{\bar n})$ & 6/0 & 2/0 & 2/0 & 0/0 & 2/0 & 0/0 &
3/1 & 2/1 & 2/1 & 2/0 & 2/1 & 3/1 & 3/0 & 3/0 & 2/1 & 1/2 &
0/1 & 0/1 \\
$\alpha {\hat O}({\bar i}jk{\bar l}mn)+ \beta {\hat O}(i{\bar
j}{\bar k}l{\bar m}{\bar n})$ & 6/0 & 2/0 & 2/0 & 0/0 & 2/0 & 0/0 &
2/1 & 1/1 & 1/0 & 1/0 & 1/0 & 3/1 & 3/0 & 3/0 & 3/2 & 1/1 &
0/1 & 1/1 \\
$\alpha {\hat O}({\bar i}j{\bar k}lmn)+ \beta {\hat O}(i{\bar
j}k{\bar l}{\bar m}{\bar n})$ & 6/0 & 2/0 & 2/0 & 0/0 & 2/0 & 0/0 &
1/1 & 1/1 & 1/0 & 1/0 & 1/0 & 1/0 & 3/0 & 1/0 & 2/1 & 1/1 &
1/3 & 1/3 \\
$\alpha {\hat O}({\bar i}{\bar j}klmn)+ \beta {\hat O}(ij{\bar
k}{\bar l}{\bar m}{\bar n})$ & 6/0 & 2/0 & 2/0 & 0/0 & 0/0 & 0/0 &
1/0 & 1/0 & 1/0 & 1/0 & 1/0 & 0/0 & 0/0 & 3/0 & 1/1 & 0/1 & 0/2 &
0/1

\end{tabular}
\end{ruledtabular}

\end{table*}

As with the (2,$q$)-positivity conditions for $q \le 5$, the
(2,6)-positivity conditions are generated from Eq.~(\ref{eq:O2}) by
considering all ${\hat C}_{i}$ operators of degree less than or
equal to six.  Six classes of lifting (2,6)-positivity conditions
arise from lifting the six classes of (2,5)-positivity conditions.
While not shown explicitly, the representative conditions can be
readily constructed from the conditions in Table~IV.  There are also
18~classes of pure (2,6)-positivity conditions.  The ${\hat C}$
operators of these 18 conditions are given by
\begin{equation}
{\hat C}(ijklmn) = \sum_{ijklmn}{ b_{i} d_{j} e_{k} f_{l} g_{m}
h_{n} {\hat a}^{\dagger}_{i} {\hat a}^{\dagger}_{j} {\hat
a}^{\dagger}_{k} {\hat a}^{\dagger}_{l} {\hat a}^{\dagger}_{m} {\hat
a}^{\dagger}_{n}}
\end{equation}
where the $b_{i}$, $d_{i}$, $e_{i}$, $f_{i}$, $g_{i}$, $h_{i}$, and
${\hat a}^{\dagger}_{i}$ become $b_{i}^{*}$, $d_{i}^{*}$,
$e_{i}^{*}$, $f_{i}^{*}$, $g_{i}^{*}$, $h_{i}^{*}$, and $a_{i}$ when
$i={\bar i}$.  Table~V provides a representative operator for each
of the 18 classes.  Each representative operator arises from the
conic combination of potentially $2^{6}$ (or 64) six-particle
operators, which are distinguished from each other by the switching
between creation and annihilation operators.  These 64 operators are
grouped in 32 particle-hole pairs given in the rows of Table~V.  For
each of the 18 representative conditions, the nonnegative integer
weights $\alpha$ and $\beta$ of the operators in each pair are
reported. The conic combination of all 32 pairs with the weights in
the $x^{\rm th}$ column generates a representative operator for
class~$x$.  The operator for each class depends only on the 2-RDO
with the dependence on the 3-, 4-, 5-, and 6-RDOs canceling through
the conic combination.  The trace of each representative operator
against the 2-RDM generates a representative condition on the 2-RDM.
Additional (2,6)-positivity conditions can be generated from the
representative conditions through a combination of switching and
reordering of the creation and annihilation operators. From the
particle-hole pairing it is easy to observe that only one class of
the (2,6)-conditions---class 4---has particle-hole symmetry, that is
$\alpha=\beta$ in all pairs.

The (2,6)-positivity conditions yield all classes of the classical
(diagonal) $N$-representability conditions when the expansion
coefficients $b_{i}$, $d_{j}$, $e_{k}$, $f_{l}$, $g_{m}$, and
$h_{n}$ are chosen to be orthogonal unit vectors.  Classically, all
classes of (2,$q$)-conditions for $q \le 5$ are in the form of
hypermetric inequalities~\cite{Cuts,MD72}.  When $q=6$, however, new
classes of classical $N$-representability conditions
emerge~\cite{E78,MD72, Cuts,G89}. In the classical limit, the first
6 classes of pure (2,6)-positivity conditions in Table~V reduce to
hypermetric inequalities while the remaining 12 can be grouped into
cycle, parachute, and Grishukhin inequalities~\cite{G89}.

\section{Discussion and Conclusions}


Both new and known $N$-representability conditions on the 2-RDM have
been derived from the constructive solution to the
$N$-representability problem~\cite{M12b}.  In addition to all of the
previously known conditions, we generate new (2,3)-, (2,4)-, (2,5),
and (2,6)-conditions where the first number $p$ in each pair
indicates the highest $p$-RDM required to evaluate the condition
(the 2-RDM in our case) and the second number $q$ indicates the
highest RDMs canceled by conic (linear nonnegative) combinations in
the derivation of the condition. There are two classes of
(2,3)-conditions: (i) lifting conditions that are derivable from
lifting the D, Q, and G (2-positivity) conditions to the
three-particle space, and (ii) pure conditions that are not
derivable from lifting and hence, are without precedent in the
2-positivity conditions.  The (2,4)-conditions have two classes of
lifting conditions and one class of pure conditions, the
(2,5)-conditions have three classes of lifting conditions and three
classes of pure conditions, and the (2,6)-conditions have six
classes of lifting conditions and eighteen classes of pure
conditions. A similar procedure of using conic combinations to
cancel operators higher than two-body can be followed for deriving
the $(2,q)$-conditions for $q>6$.

The classical (diagonal) $N$-representability conditions~\cite{WW67,
YK69,MD72,E78,Cuts} are constraints on the two-electron reduced
density function, the diagonal part of the 2-RDM, to ensure that it
represents an $N$-electron density function.  A solution to the
diagonal problem was developed in the context of both the Boole 0-1
programming problem and the maximum cut problem of graph
theory~\cite{Cuts,P89}. The recent constructive solution of the
$N$-representability problem for fermionic density matrices extends
the classical solution to the more general quantum case. All of the
quantum conditions can be cast in the form of restricting the trace
of two-body operators (model Hamiltonians) against the 2-RDM to be
nonnegative.  In the limit that all tensors in the model
Hamiltonians are decomposed into products of orthogonal rank-one
(one-index) tensors, the quantum conditions reduce to the classical
(diagonal) conditions for all unitary transformations of the
one-electron basis set.  The quantum (2,6)-conditions presented here
reduce in the classical limit to the complete set of classical
(2,6)-conditions~\cite{MD72,Cuts}, which were shown to be complete
by Grishukhin~\cite{G89}.

A significant difference between the classical and quantum
conditions is the orthogonality (classical) or non-orthogonality
(quantum) of the rank-one tensors.  Consequently, in the classical
case the creation and annihilation operators form a commutative
algebra while in the quantum case they form a non-commutative
algebra.  The non-orthogonality leads to active $N$-representability
conditions on the 2-RDM that lack a classical analogue.  For
example, all classes of lifting conditions that we presented are
inactive in the classical limit.  Because the creation and
annihilation operators commute, each class of classical
$(2,q)$-lifting conditions reduces to a class of classical
$(2,p)$-pure conditions where $p<q$.  Furthermore, typically more
than one pure quantum condition reduces to each classical condition
in the classical limit.  Table~III shows four pure (2,4)-conditions
that reduce to the same classical condition.  These quantum
conditions differ only in the ordering of the creation and
annihilation operators---a difference that disappears in the
classical, commutative limit.

The conic combination of the extreme two-body operators in the
$N$-representability conditions forms a convex set (cone) of model
Hamiltonians for which the $N$-representability conditions are
exact.  From the perspective of quantum information the
computational complexity of enforcing all $N$-representability
conditions on the 2-RDM can be shown to be non-deterministic
polynomial-time (NP) complete, meaning that in the worst-case
scenario enforcing exact $N$-representability scales
non-polynomially with system size.  Despite this complexity,
however, many realistic quantum systems are much more tractable than
the worst-case scenario implies. For example, the 2-positivity
conditions, particularly the G condition, are exact for pairing
Hamiltonians whose ground states are antisymmetrized geminal power
wavefunctions.  Such pairing Hamiltonians have been employed to
model the Cooper pairing and long-range order associated with
superconductivity.  For any strength of interaction the ground-state
energy for this class of Hamiltonians can be computed in polynomial
time.

More generally, for fixed $q$ the $(2,q)$-positivity conditions,
which contain the lower positivity conditions, cover a large class
of model Hamiltonians whose ground states are computable in
polynomial time---in a time that scales polynomially with system
size.  Even when the Hamiltonian of interest is not rigorously
contained in this class, the associated $N$-representability
conditions, which intrinsically are not constrained by the
approximations of perturbation theory, may produce an accurate lower
bound on the ground-state energy.  Computational experience with the
variational calculation of the 2-RDM in atoms and
molecules~\cite{M12a,M04,P04,M06,GM08,GM10,SGM10,PGG11} shows that
sufficiently accurate lower-bound ground-state energies are often
produced with $(2,q)$-positivity conditions where $q \le 3$.

The practical implementation of the variational 2-RDM method
requires that the energy be minimized as a functional of the 2-RDM
constrained by its $N$-representability conditions.  Both the
2-positivity conditions and the T1 and T2 conditions can be
expressed as positive semidefinite constraints (also known as linear
matrix inequalities) in which metric matrices are constrained to be
positive semidefinite.  These constraints on the 2-RDM can be
imposed during the minimization of the ground-state energy through a
genre of constrained optimization known as semidefinite
programming~\cite{M04,P04,C06,FNY07,M07,A09,M11,BP12,E79}.  The
remaining $(2,q)$-positivity conditions, however, cannot be
expressed as a traditional semidefinite constraint because the
coefficients in the ${\hat C}_{i}$ operators must be tensor
decomposed to remove the dependence of the constraints on the higher
RDMs.  Practically, as described in section~\ref{sec:sd}, these
constraints can be added to the semidefinite program through
recursively generated linear inequalities, similar to those
described in Ref.~\cite{JSM07} for T2.

The constructive solution of $N$-representability establishes 2-RDM
theory as a fundamental theory for many-particle quantum mechanics
for particles with pairwise interactions.  Lower bounds on the
ground-state energy can be computed and improved systematically
within the theory. While not all of the 2-RDM conditions will be
imposed in practical calculations, a complete knowledge of the
conditions---their form and function---can be invaluable in devising
and testing approximate $N$-representability conditions for
different types of quantum systems and interactions. Like Feynman
diagrams the positivity conditions represent different physical
interactions of the electrons.  Adding positivity conditions to the
2-RDM calculation expands the class of exactly describable model
Hamiltonians.  Just as classes of Feynman diagrams differ in
importance according to the nature of the interaction, for a given
system some positivity conditions will be significantly more
important than others.  For example, both the G and T2 conditions
have proven to be especially important in calculations of
many-electron atoms and molecules~\cite{M04,P04,M05} while the T1
condition has rarely been of any significance.  Similar evaluations
must be performed in a variety of many-electron quantum systems for
the conditions resulting from the constructive solution.

Previous variational 2-RDM computations on metallic hydrogen
chains~\cite{SGM10}, polyaromatic hydrocarbons~\cite{GM08,PGG11},
and firefly luciferin~\cite{GM10} show that they can capture strong,
multi-reference correlation effects for which appropriate
ans{\"a}tze for the wavefunction are difficult to construct. With a
suitable choice of $N$-representability conditions, therefore,
strong electron correlation effects can be computed at a
computational cost that scales polynomially with the system size.
Although the exploration of the conditions following from the
constructive solution is still in its earliest stages, a 2-RDM-based
theory with systematically improvable accuracy promises fresh
theoretical and computational possibilities for treating strong
correlation in quantum many-electron systems.

\begin{acknowledgments}

The author thanks D. Herschbach, H. Rabitz, and A. Mazziotti for
encouragement, and the NSF, ARO, Microsoft Corporation, Dreyfus
Foundation, and David-Lucile Packard Foundation for support.

\end{acknowledgments}

\bibliography{SFDM5BIBR2}

\end{document}